# Friction and Radiative Heat Exchange in a System of Two Parallel Plates Moving Sideways : Levin-Polevoy-Rytov Theory Revisited

G.V. Dedkov and A.A. Kyasov

Nanoscale Physics Group, Kabardino-Balkarian State University, Nalchik, 360004 Russia

It is shown that the fundamental results obtained in the works by Levine, Polevoi, Rytov (1980) and Polevoi (1990), based on the fluctuation-electromagnetic theory by Levine and Rytov, adequately describe the rate of radiative heat exchange and the frictional force in a system of two parallel thick plates in relative lateral motion. A numerically calculated friction force for good metals and thin gaps turns out to be by a factor $10^7$ higher than earlier obtained by Polevoi and increases with increasing conductivity of the metals.

## 1. Introduction

Fluctuation-electromagnetic theory formulated by Levine and Rytov [1] is an extension of Rytov's theory [2]. In [1], the spectrum of electromagnetic fluctuations of a heated body at an arbitrary distance from the surface of the body is expressed through mixed losses of the two point-dipole sources nearby it, which are calculated from the solution of regular electrodynamic problem. This is the essence of the generalized Kirchhoff's law, which represents the form of a fluctuation-dissipation theorem.

Within the framework of theory [1], expressions for the rate of heat exchange between the two semi-infinite media (thick plates) separated by a vacuum gap of finite width were obtained [3, 4], as well as the dissipative frictional force, arising in the case of relative lateral motion of one of the plates [5]. The first calculation of radiation heat exchange (within the framework of theory [2]) between the two plates in rest was carried out by Polder and van Hove [6] assuming a simpler case of two identical plates and a small temperature difference between them. In contrast to this, in [3, 4] media 1 and 2 were assumed to be homogeneous and isotropic with permittivities and permeabilities $\varepsilon_1, \mu_1$ and $\varepsilon_2, \mu_2$, being the complex functions of the frequency $\omega$. Moreover, the general case of anisotropic media was also examined. Later, similar problems were solving by a number of authors, but the formula for the thermal energy flux was reproduced in many cases either without references to [3, 4] (see, for example, [7, 8]) or in another equivalent form [9, 10].

The situation with work [5] turned out to be much more dramatic: in the final formulas for dissipative frictional force in the linear velocity approximation, reported by Polevoi [5], the dependence $F \propto V/c^3$ has appeared ($c$ is the speed of light in vacuum), whereas later several authors obtained linear in $V$ and independent of $c$ expressions for this force (at a finite temperature of the plates) [11, 12], or dependence $F \sim V^3$ in the quantum zero-temperature limit [13]. These contradictions "poured fuel to the fire" of a lengthy discussion on the magnitude of the dissipative force, which has been started even earlier [14] and is being not completed so far [10-13, 15-17] (for many other references see in [18, 19]).

In this work we show that the basic results for the friction force obtained in [3-5] are completely consistent with all results obtained by other authors later, while the dependence $F \propto V/c^3$ and a very low numerical value of the frictional stress (about $10^{-14} N/m^2$ for metallic plates at room temperature, at a gap width of 10 $nm$ and a relative velocity of $1 m/s$), is due to a special form of material properties of interacting bodies. We have recalculated numerically the frictional force for good nonmagnetic metals using the dielectric permittivity $\varepsilon(\omega) = i4\pi\sigma/\omega$ ($\sigma$ – the conductivity) and obtained much higher values (by $10^7$ times) in the case of thin gaps. Moreover, a striking fact is that the friction force between the metallic plates increases with increasing conductivity.

## 2. Problem statement and general expression for the tangential force by Polevoi

In a system configuration used by Polevoi [5], the Cartesian laboratory coordinate system fixed in (plate) 1 (Fig.1) is chosen so that the axis $z = x_3$ is orthogonal to the boundaries of the plates, the axis $x = x_1$, without loss of generality, is parallel to the velocity $\mathbf{V}$ of plate 2, the axis $y = x_2$ (not shown in Fig.1) is orthogonal to the $x$ and $z$ axes. The temperatures of the plates are held constant at $T_1$ and $T_2$, respectively.

Following [5], the resulting force densities $\mathbf{F}_1$ and $\mathbf{F}_2$ acting on a unit surface area of plates 1 and 2 differ only in sign: $\mathbf{F}_1 = -\mathbf{F}_2 = -\mathbf{V}F/V$, where $F$ is the modulus of the dissipative tangential force $F$ per unit area of the moving plate 2 in the laboratory reference system associated with resting plate 1. It is expressed in terms of the heat fluxes $P_1$ and $P_2$ from plates 1 and 2 (per unit area), the flow $P_1$ being calculated through the Poynting vector in the reference frame of plate 1, and the flow $P_2$ from plate 2 is calculated in its rest system:

$$F = \frac{1}{V}\left(P_1 + P_2/\gamma\right), \qquad (1)$$

where $\gamma = 1/(1-u^2)^{1/2}$, $u = V/c$. The heat fluxes $P_1$ and $P_2$ are given by

$$P_1 = \frac{\hbar}{8\pi^3} \int_{-\infty}^{\infty} d\omega \int d^2k \left( \frac{\omega}{|\omega|} - \frac{\tilde{\omega}}{|\tilde{\omega}|} \right) \omega M(\omega, \mathbf{k}, \mathbf{u}) + \\ + \frac{1}{4\pi^3} \int_{-\infty}^{\infty} d\omega \int d^2k \left[ \frac{\Pi(T_1, \omega)}{\omega} - \frac{\Pi(T_2, \tilde{\omega})}{\tilde{\omega}} \right] \omega M(\omega, \mathbf{k}, \mathbf{u}) \tag{2}$$

$$P_2 = -\frac{\hbar}{8\pi^3} \int_{-\infty}^{\infty} d\omega \int d^2k \left( \frac{\omega}{|\omega|} - \frac{\tilde{\omega}}{|\tilde{\omega}|} \right) \tilde{\omega} M(\omega, \mathbf{k}, \mathbf{u}) - \\ - \frac{1}{4\pi^3} \int_{-\infty}^{\infty} d\omega \int d^2k \left[ \frac{\Pi(T_1, \omega)}{\omega} - \frac{\Pi(T_2, \tilde{\omega})}{\tilde{\omega}} \right] \tilde{\omega} M(\omega, \mathbf{k}, \mathbf{u}) \tag{3}$$

where $\hbar$ – the Planck constant, $\mathbf{k} = (k_1, k_2)$ – a two-dimensional wave vector coplanar to the plates, $\tilde{\omega} = \gamma(\omega - \mathbf{kV})$, $|\mathbf{V}| = V$, $|\mathbf{u}| = u$, $\Pi(T, \omega) = \hbar|\omega|/(\exp(|\omega|/\omega_T) - 1)$, $\omega_T = T/\hbar$, $T$ is the temperature in energy units and the integration is performed over the entire space of wave vectors. The function $M(\omega, \mathbf{k}, \mathbf{u})$ has the form

$$M = \frac{4|q|^2}{|Q|^2} \left[ \mathrm{Im}\left(\frac{q_1}{\varepsilon_1}\right) \mathrm{Im}\left(\frac{\tilde{q}_2}{\tilde{\varepsilon}_2}\right) (1+\beta) |Q_\mu|^2 + \mathrm{Im}\left(\frac{q_1}{\mu_1}\right) \mathrm{Im}\left(\frac{\tilde{q}_2}{\tilde{\mu}_2}\right) (1+\beta) |Q_\varepsilon|^2 \right] + \\ + \frac{4|q|^2}{|Q|^2} \left[ \mathrm{Im}\left(\frac{q_1}{\varepsilon_1}\right) \mathrm{Im}\left(\frac{\tilde{q}_2}{\tilde{\mu}_2}\right) |\beta| |Q_{\mu\varepsilon}|^2 + \mathrm{Im}\left(\frac{q_1}{\mu_1}\right) \mathrm{Im}\left(\frac{\tilde{q}_2}{\tilde{\varepsilon}_2}\right) |\beta| |Q_{\varepsilon\mu}|^2 \right] \tag{4}$$

where $q_1 = \left(k^2 - (\omega/c)^2 \varepsilon_1 \mu_1\right)^{1/2}$, $q_2 = \left(k^2 - (\omega/c)^2 \varepsilon_2 \mu_2\right)^{1/2}$, $q = \left(k^2 - (\omega/c)^2\right)^{1/2}$. The branches of the square roots are chosen to satisfy $\mathrm{Re}\, q_{1,2} > 0$ and parameter $\beta$ is given by

$$\beta = \frac{\gamma^2 u^2 q^2 k_\perp^2}{k^2 \tilde{k}^2}. \tag{5}$$

Here $k_\perp^2 = \left[\mathbf{k} - \frac{\mathbf{u}(\mathbf{ku})}{u^2}\right]^2$ and $\tilde{\mathbf{k}} = \mathbf{k} + (\gamma - 1)\frac{\mathbf{u}(\mathbf{ku})}{u^2} - \gamma k \mathbf{u}$ ($\tilde{\mathbf{k}}$ – wave vector in the rest frame of plate 2), the tilde means that the corresponding quantities depending on $\omega$ and $k$ are taken at $\tilde{\omega}$ and $\tilde{k}$. The quantities $Q_\varepsilon, Q_\mu, Q_{\varepsilon\mu}, Q_{\mu\varepsilon}, Q$ in (4) are given by ($a$ is the gap width in Fig. 1)

$$Q_\varepsilon = (q + q_1/\varepsilon_1)(q + \tilde{q}_2/\tilde{\varepsilon}_2)\exp(qa) - (q - q_1/\varepsilon_1)(q - \tilde{q}_2/\tilde{\varepsilon}_2)\exp(-qa), \tag{6}$$

$$Q_\mu = (q + q_1/\mu_1)(q + \tilde{q}_2/\tilde{\mu}_2)\exp(qa) - (q - q_1/\mu_1)(q - \tilde{q}_2/\tilde{\mu}_2)\exp(-qa), \quad (7)$$

$$Q_{\varepsilon\mu} = (q + q_1/\varepsilon_1)(q + \tilde{q}_2/\tilde{\mu}_2)\exp(qa) - (q - q_1/\varepsilon_1)(q - \tilde{q}_2/\tilde{\mu}_2)\exp(-qa), \quad (8)$$

$$Q_{\mu\varepsilon} = (q + q_1/\mu_1)(q + \tilde{q}_2/\tilde{\varepsilon}_2)\exp(qa) - (q - q_1/\mu_1)(q - \tilde{q}_2/\tilde{\varepsilon}_2)\exp(-qa), \quad (9)$$

$$Q = Q_\varepsilon Q_\mu - 4\beta k^2 \tilde{\mu}^2 \left(1 - (\varepsilon_1\mu_1)^{-1}\right)\left(1 - (\tilde{\varepsilon}_2\tilde{\mu}_2)^{-1}\right). \quad (10)$$

We retained all the notation used in [5] with a single replacement $\kappa, \kappa_1, \kappa_2 \to k, k_1, k_2$.

## 3. Transformation of the general formula for $u = V/c \ll 1$

In the case $u = V/c \ll 1$, we have $\beta = 0, \tilde{k} = k$, and (4) with allowance for (10) reduces to

$$M = 4|q|^2 \left[ \text{Im}\left(\frac{q_1}{\varepsilon_1}\right) \text{Im}\left(\frac{\tilde{q}_2}{\tilde{\varepsilon}_2}\right) |Q_\varepsilon|^{-2} + \text{Im}\left(\frac{q_1}{\mu_1}\right) \text{Im}\left(\frac{\tilde{q}_2}{\tilde{\mu}_2}\right) |Q_\mu|^{-2} \right] \quad (11)$$

Using (11) and the identity

$$4\,\text{Im}\left(\frac{q_1}{\varepsilon_1}\right) \text{Im}\left(\frac{\tilde{q}_2}{\tilde{\varepsilon}_2}\right) \equiv -\left(\frac{q_1}{\varepsilon_1} - \frac{q_1^*}{\varepsilon_1^*}\right)\left(\frac{\tilde{q}_2}{\tilde{\varepsilon}_2} - \frac{\tilde{q}_2^*}{\tilde{\varepsilon}_2^*}\right), \quad (12)$$

as well the analogous identity with the permutation $\varepsilon_1 \leftrightarrow \mu_1, \tilde{\varepsilon}_2 \leftrightarrow \tilde{\mu}_2$, Eq. (1) for $P_1$ reduces to Eq. (6) in [3] for the spectral energy flux density of the thermal field (since $\tilde{q}_2 = q_2$, $\tilde{\varepsilon}_2 = \varepsilon_2$). In this case, $P_1 = -P_2$, $|P_1| = |P_2|$, and the different sign of these quantities is due to the different direction of the flow of thermal energy relative to the plates.

At $V \neq 0$, using (6), the expression for $|Q_\varepsilon|^{-2}$ can be rewritten in the form

$$|Q_\varepsilon|^{-2} = |\varepsilon_1 q + q_1|^{-2} |q\tilde{\varepsilon}_2 + \tilde{q}_2|^{-2} |1 - \Delta_{1e}\tilde{\Delta}_{2e}\exp(-2qa)|^{-2} |\varepsilon_1|^2 |\tilde{\varepsilon}_2|^2 |\exp(-2qa)| \quad (13)$$

where $\Delta_{1e} = \dfrac{\varepsilon_1 q - q_1}{\varepsilon_1 q + q_1}$, $\tilde{\Delta}_{2e} = \dfrac{\tilde{\varepsilon}_2 q - \tilde{q}_2}{\tilde{\varepsilon}_2 q - \tilde{q}_2}$. For $|Q_\mu|^{-2}$ we have the same expression as for $|Q_\varepsilon|^{-2}$ with allowance for the obvious permutations $\varepsilon_1 \leftrightarrow \mu_1, \tilde{\varepsilon}_2 \leftrightarrow \tilde{\mu}_2, \Delta_{1e} \to \Delta_{1m}, \tilde{\Delta}_{2e} \to \tilde{\Delta}_{2m}$.

The integrals in (2), (3) contains the contributions from inhomogeneous (evanescent) waves ($k > \omega/c$) and from traveling waves $k \leq \omega/c$. At $k > \omega/c$ we have $q = |q|, |\exp(-2qa)| = \exp(-2qa)$, and (12) takes the form

$$\left(\frac{q_1}{\varepsilon_1} - \frac{q_1^*}{\varepsilon_1^*}\right)\left(\frac{\tilde{q}_2}{\tilde{\varepsilon}_2} - \frac{\tilde{q}_2^*}{\tilde{\varepsilon}_2^*}\right) = -\frac{\operatorname{Im}\Delta_{1e}\operatorname{Im}\tilde{\Delta}_{2e}}{|q|^2|\varepsilon_1|^2|\tilde{\varepsilon}_2|^2}|\varepsilon_1 q + q_1|^2|\tilde{\varepsilon}_2 q + \tilde{q}_2|^2. \tag{14}$$

At $k \leq \omega/c$, correspondingly, $q = -i|q|, |\exp(-2qa)| = 1$ and

$$\left(\frac{q_1}{\varepsilon_1} - \frac{q_1^*}{\varepsilon_1^*}\right)\left(\frac{\tilde{q}_2}{\tilde{\varepsilon}_2} - \frac{\tilde{q}_2^*}{\tilde{\varepsilon}_2^*}\right) = -\frac{\left(1 - |\Delta_{1e}|^2\right)\left(1 - |\tilde{\Delta}_{2e}|^2\right)}{4|q|^2|\varepsilon_1|^2|\tilde{\varepsilon}_2|^2}|\varepsilon_1 q + q_1|^2|\tilde{\varepsilon}_2 q + \tilde{q}_2|^2 \tag{15}$$

Identities analogous to (14), (15) can be easily written with the permutations $\varepsilon_1 \leftrightarrow \mu_1, \tilde{\varepsilon}_2 \leftrightarrow \tilde{\mu}_2, \Delta_{1e} \to \Delta_{1m}, \tilde{\Delta}_{2e} \to \tilde{\Delta}_{2m}$.

Substituting (11)–(15) into (1)–(3), and taking into account the analytical properties of the function $M(\omega, \mathbf{k}, \mathbf{u})$, namely [5]

$$\begin{aligned} M(\omega, \mathbf{k}, \mathbf{u}) &> 0,\ \omega\tilde{\omega} > 0 \\ M(\omega, \mathbf{k}, \mathbf{u}) &< 0,\ \omega\tilde{\omega} < 0 \end{aligned} \tag{16}$$

we obtain the following expression for the tangential force acting on the moving plate 2 in the laboratory coordinate system (negative values of $F_x$ correspond to the dissipative frictional force)

$$\begin{aligned} F = &-\frac{\hbar}{4\pi^3}\int_0^\infty d\omega \int_{k>\omega/c} d^2k\, k_x \exp(-2qa)\operatorname{Im}\Delta_{1e}\operatorname{Im}\tilde{\Delta}_{2e}|D_e|^{-2}[\coth(\hbar\tilde{\omega}/2T_2) - \coth(\hbar\omega/2T_1)] - \\ &-\frac{\hbar}{16\pi^3}\int_0^\infty d\omega \int_{k\leq\omega/c} d^2k\, k_x \left(1 - |\Delta_{1e}|^2\right)\left(1 - |\tilde{\Delta}_{2e}|^2\right)|D_e|^{-2}[\coth(\hbar\tilde{\omega}/2T_2) - \coth(\hbar\omega/2T_1)] + (e \leftrightarrow m) \end{aligned} \tag{17}$$

where $|D_e| = |1 - \Delta_{1e}\tilde{\Delta}_{2e}\exp(-2qa)|$ and the terms $(e \leftrightarrow m)$ are determined by the same integrals by replacing $\varepsilon_1 \leftrightarrow \mu_1, \tilde{\varepsilon}_2 \leftrightarrow \tilde{\mu}_2, \Delta_{1e} \to \Delta_{1m}, \tilde{\Delta}_{2e} \to \tilde{\Delta}_{2m}$. It should be emphasized once again that in this case $\tilde{\omega} = \omega - k_x V$.

Formula (17) completely includes all the results of other authors [10–13, 17–19] at $u = V/c \ll 1$, obtained in the nonretarded and retarded limits. In particular, at $T_1 \to 0, T_2 \to 0$ from (16) one obtains the formula for the quantum frictional force between the two smooth plates [13, 17]

$$F = \frac{\hbar}{4\pi^3} \int_{-\infty}^{\infty} dk_y \int_0^{\infty} dk_x k_x \int_0^{k_x V} d\omega \exp(-2ka) \operatorname{Im}\Delta_{1e} \operatorname{Im}\tilde{\Delta}_{2e} |D_e|^{-2} + (e \leftrightarrow m). \qquad (18)$$

In turn, in the case of resting plates ($V = 0$), the formula for the resultant energy flux $P_1$ of the thermal field from plate 1 (for definiteness), taking into account (11)–(16), reduces to [8-10]

$$P_1 = \frac{\hbar}{2\pi^3}\int_0^{\infty} d\omega\,\omega \int_{k>\omega/c} d^2k \exp(-2qa) \operatorname{Im}\Delta_{1e} \operatorname{Im}\Delta_{2e} |D_e|^{-2}[n_1(\omega) - n_2(\omega)] +$$
$$+ \frac{\hbar}{8\pi^3}\int_0^{\infty} d\omega\,\omega \int_{k\le\omega/c} d^2k\left(1 - |\Delta_{1e}|^2\right)\left(1 - |\Delta_{2e}|^2\right)|D_e|^{-2}[n_1(\omega) - n_2(\omega)] + (e \leftrightarrow m), \qquad (19)$$

where $n_i(\omega) = 1/(\exp(\hbar\omega/T_i) - 1), i = 1,2$.

Another useful expression for the frictional force stems from (17) with allowance for (13)–(15):

$$F = -\frac{\hbar}{4\pi^3}\int_0^{\infty} d\omega \int d^2k\, k_x \left[\operatorname{Im}\left(\frac{q_1}{\varepsilon_1}\right)\operatorname{Im}\left(\frac{\tilde{q}_2}{\tilde{\varepsilon}_2}\right)\frac{|q|^2}{|Q_\varepsilon|^2} + \operatorname{Im}\left(\frac{q_1}{\mu_1}\right)\operatorname{Im}\left(\frac{\tilde{q}_2}{\tilde{\mu}_2}\right)\frac{|q|^2}{|Q_\mu|^2}\right].$$
$$\cdot\left[\coth\left(\frac{\hbar\tilde{\omega}}{2T_2}\right) - \coth\left(\frac{\hbar\omega}{2T_1}\right)\right] \qquad (20)$$

It is worth noting that the inner integral in (20) includes the contribution from both evanescent and traveling modes. It turns out that Eq. (20) is most convenient when calculating the frictional force between normal metals. In the same way, the expression for $P_1$ (in the case $0 < V/c \ll 1$) can be written in the form

$$P_1 = \frac{\hbar}{4\pi^3}\int_0^\infty d\omega\omega \int d^2k \left[ \mathrm{Im}\!\left(\frac{q_1}{\varepsilon_1}\right)\mathrm{Im}\!\left(\frac{\tilde{q}_2}{\tilde{\varepsilon}_2}\right)\frac{|q|^2}{|Q_\varepsilon|^2} + \mathrm{Im}\!\left(\frac{q_1}{\mu_1}\right)\mathrm{Im}\!\left(\frac{\tilde{q}_2}{\tilde{\mu}_2}\right)\frac{|q|^2}{|Q_\mu|^2} \right]$$
$$\cdot \left[\coth\!\left(\frac{\hbar\tilde{\omega}}{2T_2}\right) - \coth\!\left(\frac{\hbar\omega}{2T_1}\right)\right] \tag{21}$$

Obviously, Eq. (21) reduces to (19) at $V = 0$. Thus, the fundamental results of theory [3-5] (Eqs. (1)–(10)) fully include all the results of other authors for the rate of heat transfer and frictional force in the configuration of two arbitrary semi-infinite media (thick plates) in relative nonrelativistic motion.

## 4. Certain consequences and particular cases

In addition to Eqs. (17), (18) and (20) following from (1)–(3), it is expedient to examine some other known limits.

In the case of low sliding velocity and a rather high temperature $T_1 = T_2 = T$, $k_x V/\omega_w \ll 1$ ($\omega_w = T/\hbar$ is the Wien temperature), the temperature factor in (17) and (20) is

$$\coth(\hbar\tilde{\omega}/2T_2) - \coth(\hbar\omega/2T_1) \approx -2k_x V \frac{dn}{d\omega},\quad n(\omega) = \frac{1}{\exp(\omega/\omega_w)-1} \tag{22}$$

On the other hand, in the first-order expansion by $V$, the quantities with "tilde" in (17) and (20) will contain the velocity-independent terms, and the velocity-proportional ones. Therefore, according to (20), (22), the first-order-velocity approximation to the frictional force is given by

$$F = \frac{\hbar V}{2\pi^2}\int_0^\infty d\omega \frac{dn(\omega)}{d\omega}\int_0^\infty dk\, k^3 \left[\mathrm{Im}\!\left(\frac{q_1}{\varepsilon_1}\right)\mathrm{Im}\!\left(\frac{q_2}{\varepsilon_2}\right)\frac{|q|^2}{|Q_\varepsilon|^2} + \mathrm{Im}\!\left(\frac{q_1}{\mu_1}\right)\mathrm{Im}\!\left(\frac{q_2}{\mu_2}\right)\frac{|q|^2}{|Q_\mu|^2} \right] \tag{23}$$

It should be emphasized that all quantities in $|Q_\varepsilon|^{-2}$ and $|Q_\mu|^{-2}$ (see (13)) must be free of "tilde". In the same way, Eq. (17) reduces to

$$F = \frac{\hbar V}{2\pi^3}\int_0^\infty d\omega \frac{dn(\omega)}{d\omega}\int_{k>\omega/c} d^2k\, k_x^2 \exp(-2qa)\,\mathrm{Im}\Delta_{1e}\,\mathrm{Im}\Delta_{2e}|D_e|^{-2} +$$
$$+ \frac{\hbar V}{8\pi^3}\int_0^\infty d\omega \frac{dn(\omega)}{d\omega}\int_{k\le\omega/c} d^2k\, k_x^2 \left(1-|\Delta_{1e}|^2\right)\left(1-|\Delta_{2e}|^2\right)|D_e|^{-2} + (e \leftrightarrow m) \tag{24}$$

In the limiting case of absolutely black bodies, $\varepsilon_1 = \varepsilon_2 = \mu_1 = \mu_2 = 1$, the magnitude of the integrand in square brackets in (23) equals (–2) at $k < \omega/c$ and zero otherwise. The resultant integral yields

$$F = -\frac{\pi^2}{60}\frac{\hbar V}{c^4}\left(\frac{T}{\hbar}\right)^4 \tag{25}$$

The same result stems from (24) since $\Delta_{1e} = \Delta_{2e} = \Delta_{1m} = \Delta_{2m} = 0$. At $T = 300K$ and $V = 1m/s$ Eq. (25) yields $F \sim 5 \cdot 10^{-15} N/m^2$.

In the particular case $\omega_w a/c \ll 1$ (i. e. $a \ll 7.6\mu m$ at $T = 300K$), if use is made of the approximation $(k^2 - \omega^2/c^2)^{1/2} \approx k$, one obtains $\Delta_{ie} \approx (\varepsilon_i - 1)/(\varepsilon_i + 1)$ and $\Delta_{im} \approx \frac{\omega^2(\varepsilon_i - 1)}{4k^2 c^2}$ ($i = 1,2$). Then, for dielectrics and poor conductors, $\Delta_{im} \approx 0$, and Eq. (24) transforms to

$$F \approx \frac{\hbar V}{2\pi^2}\int_0^\infty \frac{dn(\omega)}{d\omega}d\omega \operatorname{Im}\left(\frac{\varepsilon_1-1}{\varepsilon_1+1}\right)\operatorname{Im}\left(\frac{\varepsilon_2-1}{\varepsilon_2+1}\right)\int_0^\infty dk\, k^3 \exp(-2ka)|D_e|^{-2}, \tag{26}$$

where

$$D_e = 1 - \left(\frac{\varepsilon_1-1}{\varepsilon_1+1}\right)\left(\frac{\varepsilon_2-1}{\varepsilon_2+1}\right)\exp(-2ka) \tag{27}$$

For conductors with permittivities $\varepsilon_{1,2} = 1 + i4\pi\sigma_{1,2}/\omega$, assuming $\omega_{w1,2}/2\pi\sigma_{1,2} \ll 1$, we have $|D_e| \approx 1$ and Eq. (26) yields

$$F \approx -\frac{\varsigma(3)}{32\pi^4}\frac{(T/\hbar)^2}{\sigma_1\sigma_2}\frac{\hbar V}{a^4}, \tag{28}$$

where $\varsigma(3) = 1.202$ – Riemann's zeta-function. At $T = 300K, V = 1m/s, \sigma_1 = \sigma_2 = 10^{14} s^{-1}$ (graphite), and $a = 10nm$ Eq. (28) yields $F \sim 6 \cdot 10^{-7} N/m^2$. We note that the value of $F$ decreases with increasing conductivities.

For good metals, however, the above argumentation is not valid and a more accurate calculation is required. In [5], the impedance approximation with the factor $(k^2 - \varepsilon_i\mu_i\omega^2/c^2)^{1/2} \approx i(\varepsilon_i\mu_i)^{1/2}\omega/c = i\varepsilon_i\varsigma_i$ was used ($\varsigma_i$ – the impedance). This led to the dependence $F \sim VT^{7/2}/\sigma^{1/2}a$ (at a small gap width) for good metals with $\sigma_1 = \sigma_2 = \sigma$ and

$\mu_1 = \mu_2 = 1$. The corresponding numerical assessment results in $\sim 3 \cdot 10^{-14} N/m^2$ [5] at the same conditions as above and assuming that $\sigma = 5 \cdot 10^{17} s^{-1}$. However, as we will show in what follows, a more accurate numerical calculation with the use of the exact factor $(k^2 - \varepsilon_i \mu_i \omega^2 / c^2)^{1/2}$ for good nonmagnetic metals leads to a considerably higher frictional force.

## 5. The case of good nonmagnetic metals

Let us transform Eq. (20) to a form convenient for further numerical computation. We consider the case of identical metals $\sigma_1 = \sigma_2 = \sigma$, $\mu_1 = \mu_2 = 1$, $\varepsilon(\omega) = i4\pi\sigma/\omega$. Then

$$(k^2 - \varepsilon\omega^2/c^2)^{1/2} = \left(k^4 + \frac{(4\pi\sigma\omega)^2}{c^4}\right)^{1/4} \exp(i\phi) \, , \, \phi = -0.5 \, arctg\left(\frac{4\pi\sigma\omega}{k^2 c^2}\right) \tag{29}$$

To find the contribution from evanescent waves $k > \omega/c$, we introduce the new variables $\omega = \omega_w x$ ($\omega_w = T/\hbar$) and $k^2 = (y^2 + \lambda_w^2 x^2)/a^2$ ($\lambda_w = \omega_w a/c$), $kdk = ydy/a^2$. The most important contribution is related to the second term in square brackets of (23). The corresponding inner integral transforms to

$$\int\limits_{\omega/c}^{\infty} dk k^3 \, \text{Im}\left(\frac{q_1}{\mu_1}\right) \text{Im}\left(\frac{q_2}{\mu_2}\right) \frac{|q|^2}{|Q_\mu|^2} = \frac{1}{4a^4} \int\limits_0^{\infty} dy y^3 (y^2 + \lambda_w^2 x^2) P_1^2 |Q_{\mu 1}|^{-2} \sin^2(\phi) \equiv \frac{1}{a^4} I_{\mu 1}, \tag{30}$$

$$T_1 = \left[(y^2 + \lambda_w^2 x^2)^2 + \lambda^2 x^2\right]^{1/4}, \lambda = \frac{4\pi\sigma\omega_w a^2}{c^2} \tag{31}$$

$$|Q_{\mu 1}|^2 = |(y^2 + P_1^2 \exp(2i\phi))\sinh(y) + 2yP_1 \exp(i\phi)\cosh(y)|^2 \tag{32}$$

$$\phi = -0.5 \, arctg\left(\frac{\lambda x}{y^2 + \lambda_m^2 x^2}\right) \tag{33}$$

The contribution from propagating waves $k < \omega/c$ is obtained with substitutions $\omega = \omega_w x$ and $k^2 = \frac{\omega_w^2}{c^2}(x^2 - y^2)$, $kdk = -ydy\omega_w^2/c^2$. Then we obtain

$$\int_0^{\omega/c} dk k^3 \operatorname{Im}\left(\frac{q_1}{\mu_1}\right) \operatorname{Im}\left(\frac{q_2}{\mu_2}\right) \frac{|q|^2}{|Q_\mu|^2} = \frac{\omega_w^4}{4c^4} \int_0^x dy y^3 (x^2 - y^2) P_2^2 |Q_{\mu 2}|^{-2} \sin^2(\phi) \equiv \frac{\omega_w^4}{c^4} I_{\mu 2} \qquad (34)$$

$$T_2 = \left((x^2 - y^2)^2 + \tilde{\lambda}^2 x^2\right)^{1/4}, \quad \tilde{\lambda} = \frac{4\pi\sigma}{\omega_w} \qquad (35)$$

$$|Q_{\mu 2}|^2 = \left|(y^2 - P_2^2 \exp(2i\phi)) \sin(\lambda_w y) - 2 y P_2 \exp(i\phi) \cos(\lambda_w y)\right|^2 \qquad (36)$$

$$\phi = -0.5 \operatorname{arctg}\left(\frac{\tilde{\lambda} x}{x^2 - y^2}\right) \qquad (37)$$

Making use the same substitution for variables $\omega$, $k$, the integrals corresponding to the first term in square brackets of (23) transform to:

i) evanescent-wave branch

$$\int_{\omega/c}^\infty dk k^3 \operatorname{Im}\left(\frac{q_1}{\varepsilon_1}\right) \operatorname{Im}\left(\frac{q_2}{\varepsilon_2}\right) \frac{|q|^2}{|Q_\varepsilon|^2} = \frac{1}{4a^4} \int_0^\infty dy y^3 (y^2 + \lambda_w^2 x^2) x^2 \tilde{\lambda}^{-2} P_1^2 |Q_{\varepsilon 1}|^{-2} \cos^2(\phi) \equiv \frac{1}{a^4} I_{\varepsilon 1}, \qquad (38)$$

$$|Q_{\varepsilon 1}|^2 = \left|(y^2 - x^2 \tilde{\lambda}^{-2} P_1^2 \exp(2i\phi)) \sinh(y) - i 2 x \tilde{\lambda}^{-1} y P_1 \exp(i\phi) \cosh(y)\right|^2 \qquad (39)$$

ii) propagating-wave branch

$$\int_0^{\omega/c} dk k^3 \operatorname{Im}\left(\frac{q_1}{\varepsilon_1}\right) \operatorname{Im}\left(\frac{q_2}{\varepsilon_2}\right) \frac{|q|^2}{|Q_\varepsilon|^2} = \frac{\omega_w^4}{4c^4} \int_0^x dy y^3 x^2 \tilde{\lambda}^{-2} (x^2 - y^2) P_2 |Q_{\varepsilon 2}|^{-2} \cos^2(\phi) \equiv \frac{\omega_w^4}{c^4} I_{\varepsilon 2} \qquad (40)$$

$$|Q_{\varepsilon 2}|^2 = \left|i(y^2 + x^2 \tilde{\lambda}^{-2} P_2^2 \exp(2i\phi)) \sin(\lambda_w y) - 2 x \tilde{\lambda}^{-1} y P_2 \exp(i\phi) \cos(\lambda_w y)\right|^2 \qquad (41)$$

Finally, substituting (30), (34), (38), (40) into (23) yields

$$F = -\frac{\hbar V}{2\pi^2 a^4} \int_0^\infty dx \frac{\exp(x)}{(\exp(x)-1)^2} \left(I_{\mu 1} + I_{\mu 2} \lambda_w^4 + I_{\varepsilon 1} + I_{\varepsilon 2} \lambda_w^4\right) \qquad (42)$$

As follows from (34), (40) and (42), the contributions from propagating modes in the force $F$ are independent of distance $a$, and therefore these terms will be insignificant at small gap widths. Unlike this, the contributions from evanescent modes will be negligible at large gap widths. The calculation results are shown in Figs. 2–5 (the values of $F_x$ are given with a positive sign).

Solid curves 1 and 2 in Fig. 2 correspond to the terms $I_{\mu 1}$ and $I_{\varepsilon 1}$ in (42) (i. e. the contributions from evanescent modes). The dashed curve was calculated according to [5], namely

$$F_P = -\frac{105\varsigma(7/2)}{2^{17/2}\cdot\pi}\frac{T^{7/2}}{\sigma^{1/2}\hbar^{5/2}c^3}\frac{V}{a} \qquad (43)$$

where $\varsigma(7/2) = 1.127$. The contribution from propagating modes in (42) (not shown in Fig. 2) are negligible and become more noticeable only at the distances more than $10\,\mu m$. These contributions are shown separately in Fig. 3. As one can see from Fig. 2, the frictional force for the plates of good metals proves to be $10^7$ times higher in comparison with the original calculation by Polevoi [5]. In addition, the dominating contribution stems from the magnetic terms in (41). In the range of the gap widths $1 \leq a \leq 30\ nm$ the force goes down slightly slower than $1/a$, but with a further increase of the gap width, the slope of the curve is closer to $1/a^2$. The increase of the frictional force at a small gap width is physically due to the high values of the reflection coefficients of the metal plates, as a result of which the electromagnetic waves that acquire the Doppler shift due to their relative motion are repeatedly reflected from plates, resulting in an increase in the frictional force. In this case, the magnitude of the frictional force is 7-8 orders of magnitude higher than in the case of absolutely black plates.

Another striking fact is that the frictional force increases with increasing conductivity of metals (approximately as $F_x \propto \sigma^{3/2}$). This is illustrated in Fig. 4 at various temperatures, assuming $a = 10\,nm$, $V = 1\,m/s$. Note that the relative conductivity $\sigma/\sigma(300) = 10^{-4}$ ($\sigma(300) = 5\cdot 10^{17}\ s^{-1}$) corresponds to poor conductors like graphite. In the same way, Fig. 5 shows the frictional force as a function of temperature and conductivity. From Fig. 4, 5 it follows the fallacy of the claim that the frictional force is maximal in the case of poor conductors of the type of graphite [10, 13].

It is interesting to compare the calculated values of $F_x$ with the measured dissipative force in experiment [20], corresponding to the geometry of the spherical probing tip (of gold) with a curvature radius of $1\,\mu m$, moving above a flat Au-coated mica surface: $\sim 1.5\cdot 10^{-13}\,N$ at

$a = 10\,nm$, $V = 1\,m/s$. Assuming that the tip has the cylindrical form, its end-face has a radius of $1\,\mu m$, and the gap width is 10 *nm*, the calculated force $F_x$ proves to be $\sim 2.8 \cdot 10^{-18}\,N$, i. e. it is too low to explain the results in [20]. At the same time, the distance dependence ($F \sim d^{-\alpha}$) and the temperature dependence $F(T)$ of the force prove to be close to those observed in [20]. So, according to [20], $\alpha = 1.3 \pm 0.3$ and $F(300)/F(77) \approx 6$ at $a = 20\,nm$, whereas from our calculation it follows $\alpha = 1 \div 2$ and $F(300)/F(77) = 5.8$.

**Conclusions**

We have proved that the fundamental expressions for the frictional force and the rate of radiative heat exchange between two halfspaces (thick plates) separated by a thin gap, obtained in the works by Levine, Rytov and Polevoi, are in full agreement with the works by other authors in the case of nonrelativistic relative velocities. The case of relativistic velocities needs a special consideration.

We also came to the conclusion, that in the case of good metals the frictional force is higher by a factor $10^7$ as compared to the earlier assessment by Polevoi. Though the absolute value of the frictional force is small compared to the dissipative force observed in [20], its temperature and distance dependences agree well with the experiment. Another important result is that the frictional force increases with increasing conductivity of the metals.

In our opinion, the measurement of the frictional force will be more realistic when using the tips with a radius of ~100 $\mu m$. Moreover, it would be interesting to examine the behavior of this force at temperatures close to the temperature of superconducting transition, since the growth in the conductivity can compensate its drop with decreasing temperature.

**References**

[1] M.L. Levin and S.M. Rytov, *Theory of equilibrium thermal fluctuations in electrodynamics*, Moscow, Nauka,1967.
[2] S.M. Rytov, *Theory of electric fluctuations and thermal radiation*, Moscow, Acad.Sci. USSR, 1953 (in Russian).
[3] M.L. Levin, V.G. Polevoi, and S.M. Rytov, Sov. Phys. JETP 52(6) (1980) 1054.
[4] V.G. Polevoi, *Heat exchange by fluctuating electromagnetic field*, Moscow, Nauka, 1990 (in Russian)
[5] V.G. Polevoi, Sov. Phys. JETP 71(6) (1990) 1119.
[6] D. Polder and M. Van Hove Phys. Rev. B4 (1971) 3303.

[7] J.J. Loomis and H.J. Maris, Phys. Rev. B50 (1994)18517.

[8] K. Park K. and Zhang Zhuomin, Frontiers in Heat and Mass Transfer 4 (2013) 013001.

[9] V.B.Bezerra, G. Bimonte, G.L. Klimchitskaya, V. M. Mostepanenko, and C. Romero, Eur. Phys. J. C52 (2007) 701.

[10] A.I. Volokitin, B. N. J. Persson, Rev. Mod. Phys. 79 (2007)1291.

[11] B. N. J. Persson, Zhang Zhenyu, Phys. Rev. B57 (1998) 7327.

[12] A.I. Volokitin, B. N. J. Persson, J. Phys. C. 11 (1999), 345.

[13] J. B. Pendry, J. Phys. C. 9 (1997) 10301.

[14] V.E. Teodorovich, Proc. Roy. Soc. London A. 362 (1978) 71.

[15] T. G. Philbin and U. Leonhardt, New J. Phys. 11 (2009) 03035; arXiv: 094.2148.

[16] A.I. Volokitin, B. N. J. Persson, New J. Phys. 11 (2009) 033035.

[17] J.B. Pendry, New J. Phys. 12 (2010) 033028.

[18] K. A. Milton, J.S. Hoye, and, I. Brevik, Symmetry 8 (2016) 29.

[19] G.V. Dedkov, A.A. Kyasov, Phys. Usp. 187 (2017) 559.

[20] B.C. Stipe, H.J. Mamin, T.D. Stowe, Y.W. Kenny, and D. Rugar, Phys. Lett. 87(9) (2001) 096801.

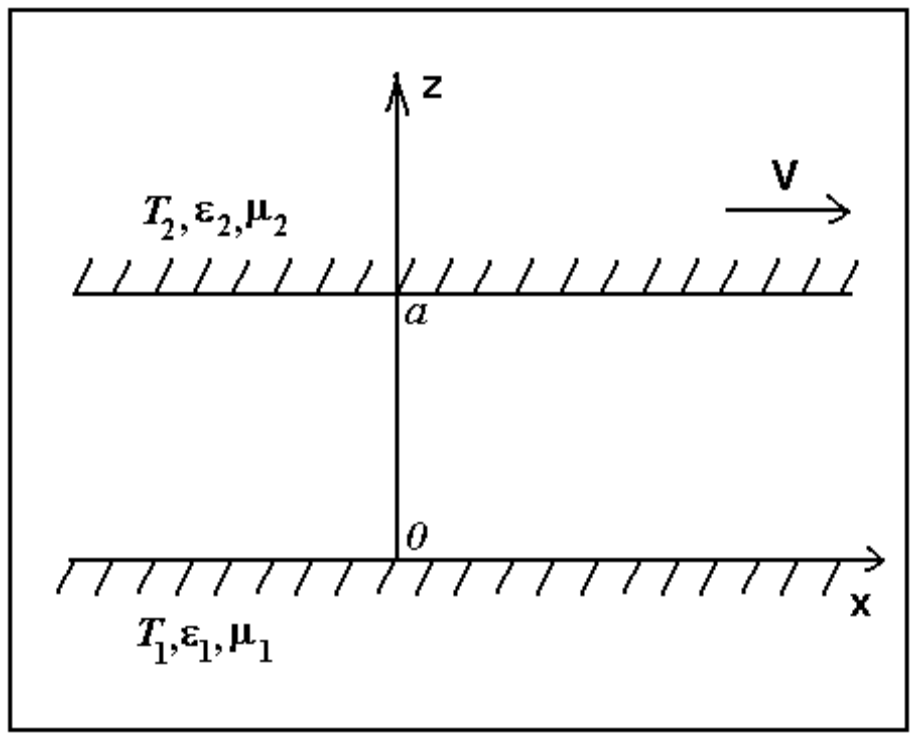

Fig. 1. Configuration of the system.

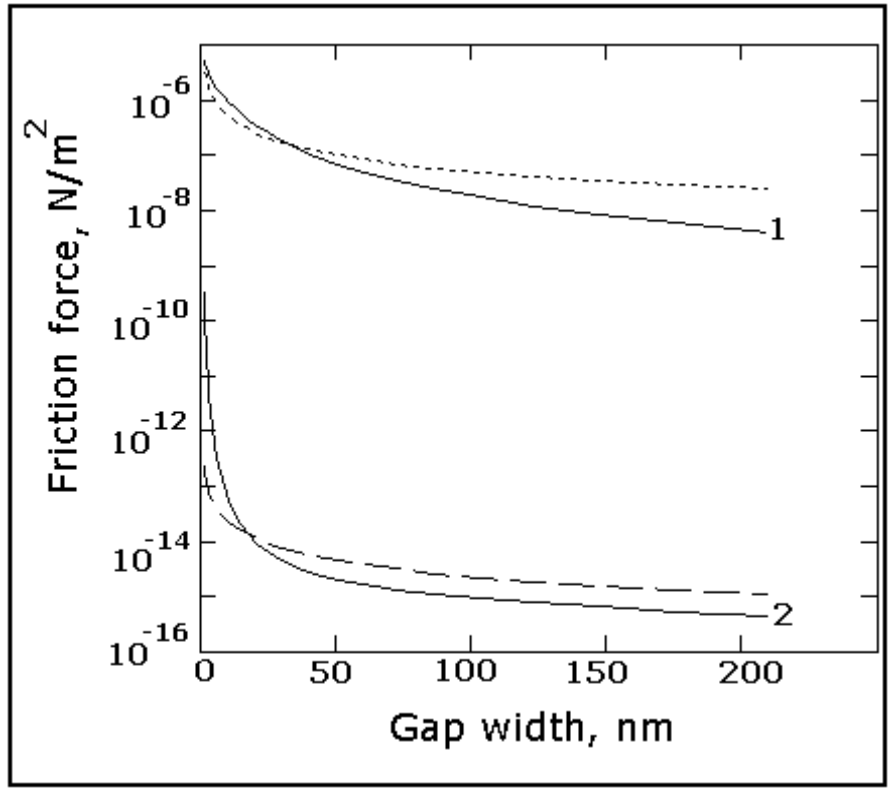

Fig.2. Frictional force per unit surface area of the vacuum contact according to (42) (solid lines ) and (43) (dashed line). Dotted line shows the fitting dependence $F \sim 1/a$, solid lines 1 and 2 correspond to the terms $I_{\mu 1}$ and $I_{\varepsilon 1}$ in (42). The used parameters are: $\sigma_1 = \sigma_2 = 5 \cdot 10^{17} \, s^{-1}, T = 300K, V = 1m/s$.

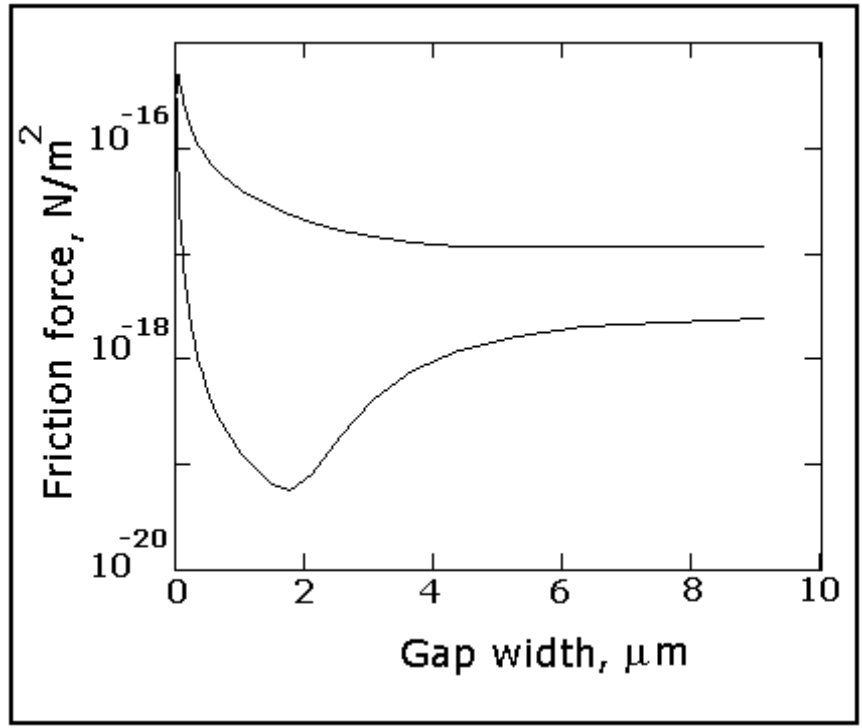

Fig. 3. Contributions to frictional force (42) from traveling modes related with the terms $I_{\varepsilon 2}$ (upper curve) and $I_{\mu 2}$ (bottom curve), $\sigma_1 = \sigma_2 = 5 \cdot 10^{17} s^{-1}, T = 300K, V = 1 m/s$.

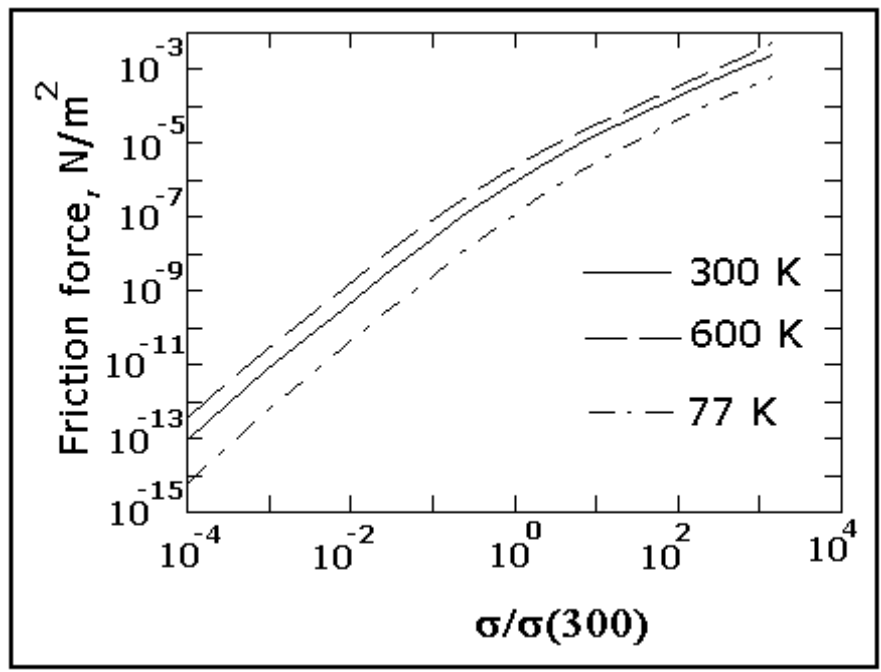

Fig.3. Frictional force as a function of conductivity and temperature. $\sigma(300) = 5 \cdot 10^{17} s^{-1}$, $a = 10 nm$

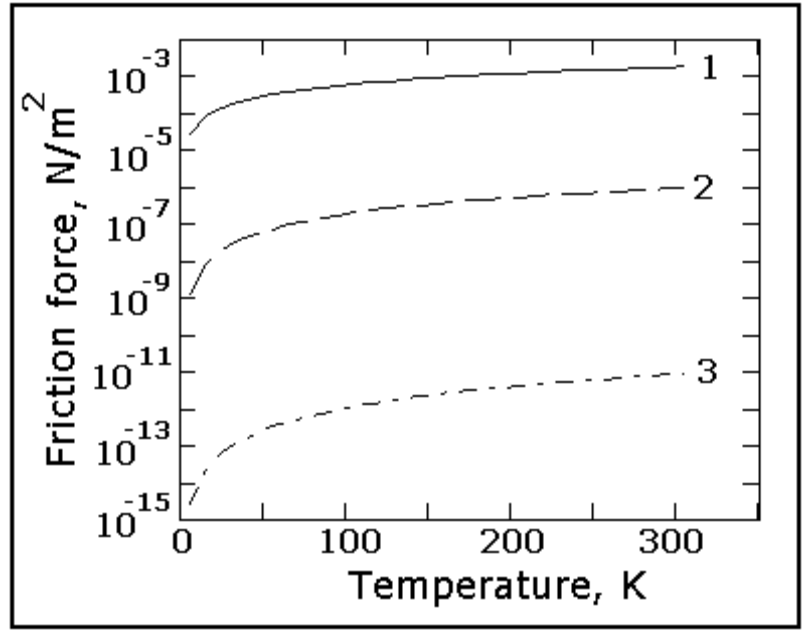

Fig. 4. Frictional force as a function of temperature and conductivity. Lines 1-3 correspond to conductivities of $5\cdot 10^{20}, 5\cdot 10^{17}$, and $5\cdot 10^{14}$ $s^{-1}$.